\documentclass[aps,
		prl,
		reprint,
		twocolumn,
		superscriptaddress,
		shortbibliography,
		nofootinbib,
		floatfix,
		notitlepage
		]{revtex4-1}
\pdfoutput=1

\usepackage{amsmath,amssymb}
\usepackage{graphicx,accents}
\usepackage[usenames,dvipsnames]{xcolor}
\usepackage{slashed}
\usepackage[normalem]{ulem}
\usepackage{bm}

\usepackage{hyperref}
\hypersetup{colorlinks=true}

\usepackage{tabularx}

\usepackage{wasysym}
\usepackage{ifsym}

\usepackage{tikz} 
\newcommand\TikCircle[1][2.5]{\tikz[baseline=-#1]{\draw[thick](0,0)circle[radius=#1mm];
\draw[thick](0,0)circle[radius=2mm];}}

\newcommand\TikPair[1][2.5]{\tikz[baseline=-#1]{
\draw[thick](0,0)--(0.3,0.25);
\draw[thick](0.075,0)--(0.3,0.175);
\draw[thick](0,0)--(0.3,-0.25);
\draw[thick](0.075,0)--(0.3,-0.175);}}

\newcommand{\expF}{\mathcal{F}}
\newcommand{\vphi}{\varphi}

\newcommand{\mbf}[1]{\mathbf{#1}}
\newcommand{\trm}[1]{\textrm{#1}}
\newcommand{\tsf}[1]{\textsf{#1}}

\newcommand{\be}{\begin{equation}}
\newcommand{\ee}{\end{equation}}
\newcommand{\bea}{\begin{eqnarray}}
\newcommand{\eea}{\end{eqnarray}}
\newcommand{\bi}{\begin{itemize}}
\newcommand{\ei}{\end{itemize}}

\newcommand{\nn}{\nonumber}

\newcommand{\prob}{\tsf{P}}

\newcommand{\ket}[1]{|#1\rangle}

\newcommand{\Supplementary}{{Appendix }}

\newcommand{\figref}[1]{Fig. \ref{#1}}

\newcommand{\eqnref}[1]{Eq. (\ref{#1})}
\newcommand{\eqnrefs}[2]{Eqs. (\ref{#1}) and (\ref{#2})}

\newcommand{\vkap}{\varkappa}

\newcommand{\ud}{\mathrm{d}}
\newcommand{\LCperp}{{\scriptscriptstyle \perp}}
\newcommand{\hilb}[1]{{\mathcal{H}}[#1]}

\makeatletter
\def\ps@pprintTitle{%
 \let\@oddhead\@empty
 \let\@evenhead\@empty
 \def\@oddfoot{}%
 \let\@evenfoot\@oddfoot}
\makeatother

\begin{document}

\title{Using nonlinear Breit-Wheeler to test nonlinear vacuum birefringence}

\author{O.~Borysov}
\affiliation{Deutsches Elektronen-Synchrotron DESY, Notkestr. 85, 22607 Hamburg, Germany}
\author{B.~Heinemann}
\affiliation{Deutsches Elektronen-Synchrotron DESY, Notkestr. 85, 22607 Hamburg, Germany}
\affiliation{Albert-Ludwigs-Universit\"at Freiburg, 79104 Freiburg, Germany}
\author{A.~Ilderton}
\affiliation{Higgs Centre, School of Physics \& Astronomy, University of Edinburgh, UK}
\author{B.~King}
\email{b.king@plymouth.ac.uk}
\affiliation{Deutsches Elektronen-Synchrotron DESY, Notkestr. 85, 22607 Hamburg, Germany}
\affiliation{Centre for Mathematical Sciences, University of Plymouth, Plymouth, PL4 8AA, United
Kingdom}
\author{A.~Potylitsyn}
\affiliation{Deutsches Elektronen-Synchrotron DESY, Notkestr. 85, 22607 Hamburg, Germany}


\date{\today}
\begin{abstract}
Quantum electrodynamics predicts that the quantum vacuum is birefringent, but due to the very small cross-section this is yet to be confirmed by experiment. Vacuum birefringence arises as the elastic part of photon-photon scattering; the inelastic part is Breit-Wheeler pair-production. We outline how measurements of the photon-polarised nonlinear Breit-Wheeler process can be used to infer a measurement of nonlinear vacuum birefringence. As an example scenario, we calculate the accuracy of such a measurement for parameters anticipated at upcoming laser-particle experiments. 
\end{abstract}
\maketitle
\twocolumngrid
The quantum vacuum, exposed to strong electromagnetic fields, can behave as a refractive medium. The propagation of probe photons then becomes polarisation-dependent, a phenomenon named vacuum birefringence. Here `vacuum' emphasises that it is \emph{virtual} electron-positron pairs which affect birefringence, as opposed to distributions of real matter in e.g.~optics. Vacuum birefringence is a manifestion of \emph{polarised} photon-photon scattering first predicted in the 1930s~\cite{Halpern:1933dya}, but remaining unobserved. \emph{Linear, unpolarised} scattering has been observed in {ultra-peripheral heavy ion collisions at ATLAS~\cite{ATLAS:2017fur,ATLAS:2019azn} and CMS~\cite{CMS:2018erd}, and} higher order effects observed in unpolarised Delbr\"uck scattering~\cite{Jarlskog:1973aui,Schumacher:1975kv}. There have been many suggestions for how to measure polarised scattering via collisions of intense laser pulses~\cite{King:2015tba,Fedotov:2022ely}, and for how to measure vacuum birefringence using X-ray photons to probe intense optical lasers~\cite{Heinzl:2006xc,DiPiazza:2006pr}, which is the focus of the planned HIBEF experiment~\cite{Schlenvoigt:2016jrd}.  Such experiments require sensitive X-ray polarimetry and face a significant challenge in separating signal from background. Suggestions for how to counter this, using e.g.~shaped beams~\cite{Karbstein:2015xra}, has seen much attention in recent years.

Here we propose a way to overcome these challenges and measure vacuum birefringence indirectly, via experiments on the, at first sight, very different process of pair production from polarised photons colliding with an intense laser, or `nonlinear Breit-Wheeler' (NBW)~\cite{Ritus1985}. NBW is the target of several upcoming experiments~\cite{Abramowicz:2021zja,Chen:22}; the unpolarised process has {so far} only been observed in the E144  experiment~\cite{Burke:1997ew,Bamber:1999zt} {as part of the two-step trident process}, while the polarised \emph{linear} process was recently measured by the STAR collaboration in ultra-peripheral heavy-ion collisions~\cite{STAR:2019wlg}. (See also~\cite{Mignani:2016fwz,Capparelli:2017mlv} for discussions of vacuum birefringence in the emissions of strongly magnetised neutron stars.)

Our proposal exploits two fundamental properties of quantum field theory. First, unitarity (the optical theorem) relates the number of produced pairs to the imaginary part of the photon forward scattering amplitude. Second, analyticity (Kramers-Kronig relations, {routinely used in nonlinear optics \cite{Hutchings1992KramersKrnigRI}}) dictates the real part of the amplitude given the imaginary part. Thus, as we make precise below, a measurement of pair yield in polarised NBW implies a measurement of polarised photon-photon scattering, and thereby vacuum birefringence. There are several advantages of such a scheme over matterless vacuum birefringence experiments; {in high-energy laser-particle experiments,} the
pair production cross-section is orders of magnitude larger than that for elastic photon-photon scattering, and positrons are easier to measure than photons within a photon background, circumventing the ``signal/noise" problem.  As an example set-up, we will consider the LUXE experiment~\cite{Abramowicz:2021zja}, employing a diamond crystal radiator to produce coherent bremsstrahlung (CB) -- as has been demonstrated at similar photon energies in e.g.~GlueX~\cite{GlueX:2020idb} -- to probe nonlinear Breit-Wheeler in the regime where the cross-section is largest. This means we can probe photon-photon scattering beyond the linear regime previously investigated~\cite{STAR:2019wlg}.

\emph{Theory}. Consider the collision of a high-energy photon and an intense laser pulse. The probability of pair creation by the photon is related, via the optical theorem, to the imaginary part of the photon forward scattering amplitude. Working to leading order in the fine-structure constant, $\alpha$, but treating the {interaction with the} intense laser exactly, the optical theorem can be expressed as
\bea 
2\,\tsf{Im}\phantom{o}^{j}\raisebox{0.4em}{\uwave{\hspace{0.5cm}}}\TikCircle\raisebox{0.4em}{\uwave{\hspace{0.5cm}}}\!\!\phantom{o}^{j} = \Big|\!\!\phantom{o}^{j}\,\raisebox{0.5em}{\uwave{\hspace{0.5cm}}}\TikPair\raisebox{0.4em}\,\Big|^{2} \label{eqn:diag0}
\eea
in which $j$ represents the state of the photon and the double line represents `dressed' electrons/positrons interacting with the intense laser. Now, for every complex function $F(z)$ (analytic in the upper-half $z$-plane and vanishing faster than $1/|z|$ as $|z|\to \infty$) its real $F^{r}$ and imaginary $F^{i}$ parts are related by $F^{r}(z) = \hilb{F^{i};z}$ in which $\hilb{F^{i};z} = \pi^{-1} {\mathrm{PV}}\int \mathrm{d}z'\,F^{i}(z')/(z'-z)$ is the Hilbert transform. Thus, if $z$ represents some appropriate variable in the NBW probability, then the Hilbert transform w.r.t.~$z$ gives the real part of the photon forward scattering amplitude as, schematically
\be
2\,\tsf{Re}\phantom{o}^{j}\raisebox{0.4em}{\uwave{\hspace{0.5cm}}}\TikCircle\raisebox{0.4em}{\uwave{\hspace{0.5cm}}}\!\!\phantom{o}^{j} = {\mathcal{H}}\left[
\!\Big|\!\!\phantom{o}^{j}\,\raisebox{0.5em}{\uwave{\hspace{0.5cm}}}\TikPair\raisebox{0.4em}\,\Big|^{2}\right] .\label{eqn:diag01}
\ee
Combining (\ref{eqn:diag0}) and (\ref{eqn:diag01}), we obtain the full one-loop amplitude from tree-level NBW. 
This is related to work by Toll~\cite{Toll:1952rq}, who applied Kramers-Kronig to the vacuum `refractive indices' (see below); however, we need to extend these formal ideas to make them experimentally relevant. In particular we need to identify suitable $F$ and $z$, and we need to know how to perform the Hilbert transformation in $z$ given some experimental data on $F$. To achieve this we exploit the behaviour of physical observables in the parameter regime of interest.

To produce a detectable number of pairs via NBW, one ideally requires the strong-field parameter of the photon $\chi$, to satisfy $\chi\gtrsim 1$ where $\chi = |e|\sqrt{-(k\cdot \mathcal{F})^{2}}/m^{3}$ ($m$ and $e<0$ are the electron mass and charge, $k$ is the photon momentum, {$\mathcal{F}_{\mu\nu}$} the Faraday tensor of the laser and we set $\hbar=c=1$). We also work in the intensity regime for which the locally constant field approximation applies~\cite{DiPiazza:2017raw,Ilderton:2018nws} -- this means the NBW probability in a focussed laser pulse can be obtained by {integrating} a local rate, calculated in a constant crossed field, with the pulse profile~\cite{King:2019igt,Blackburn:2021cuq}. This requires that the laser intensity parameter $\xi$ satisfies $\xi\gg1$ where $\xi := \chi/\eta$ and $\eta = k\cdot \vkap / m^{2}$ with $\vkap$ the characteristic wavevector of the laser. In this regime, the polarised NBW rate depends non-trivially only on the strong-field parameter $\chi$: our goal therefore is to apply the Hilbert transform in $\chi$. (Indeed it has been shown that NBW at small $\chi$ can be Hilbert-transformed to yield a resummation of the small-$\chi$ expansion of the real part of the vacuum refractive index \cite{Heinzl:2006pn}.) 

Now, let $\mathcal{M}_{ij}$ be the amplitude for a photon to scatter from polarisation state $\ket{i}$ to $\ket{j}$. If $j=1,2$ represent a basis of linear polarisations transverse to the laser propagation direction, then the optical theorem relates the amplitudes $\mathcal{M}_{jj}$ to the probability $\tsf{P}_j$ of NBW from a photon in polarisation state $\ket{j}$. Furthermore in the regime of interest the helicity flip amplitude $\mathcal{M}_{+-}$ (where $\ket{\pm} = (\ket{1} \pm i \ket{2})/\sqrt{2}$) obeys
\bea 
2\mathcal{M}_{+-} = \mathcal{M}_{11}-\mathcal{M}_{22} \label{eqn:AmpFlip}
\eea
since $\ket{1}$ and $\ket{2}$ are polarisation eigenstates and $\mathcal{M}_{12}=\mathcal{M}_{21}=0$. 
Thus, the optical theorem applied to ${\tsf{P}_1 -\tsf{P}_2}$ {is proportional} to the imaginary part of $\mathcal{M}_{+-}$, while the Hilbert transform is proportional to the real part:
\be
4\tsf{Re}\phantom{o}^{+}\raisebox{0.4em}{\uwave{\hspace{0.5cm}}}\TikCircle\raisebox{0.4em}{\uwave{\hspace{0.5cm}}}\!\!\!\!\phantom{o}^{-} = 
{\mathcal{H}}\left[
\!\Big|\!\!\phantom{o}^{1}\,\raisebox{0.5em}{\uwave{\hspace{0.5cm}}}\TikPair\raisebox{0.4em}\,\Big|^{2}\right]-{\mathcal{H}}\left[
\!\Big|\!\!\phantom{o}^{2}\,\raisebox{0.5em}{\uwave{\hspace{0.5cm}}}\TikPair\raisebox{0.4em} \,\Big|^{2}\right]\nonumber 
\ee
Thus our Hilbert transform scheme gives access to `flip' and `no flip' scattering amplitudes. (The kinematics of forward scattering yields a simple relationship between  amplitudes and probabilities $\prob_{ij}$ for the photon to change state from $\ket{i}$ to $\ket{j}$,  namely $\prob_{ij} = (\alpha/\eta)^{2} | \mathcal{M}_{ij} |^{2}$.)

\emph{Toward experiment.} Now, probe photon distributions in experiment are typically broadband, implying many different values of $\chi$ impact the pair yield. Therefore, the Hilbert transform will be over the maximum value of $\chi$, which we denote $\chi_{0}$ i.e. $\chi\in[0,\chi_{0}]$. We note in particular that it is generally easier to repeat the experiment at different $\chi_{0}$ than at e.g.~probe photon energy, because the former can be achieved simply by defocussing the laser, while the latter is determined by the photon source. Performing the Hilbert transform in $\chi_{0}$ thus allows it to be evaluated with more experimental data points, providing a more accurate inference of vacuum birefringence.

We now make these ideas precise, beginning with an estimate for the number of pairs
$N_{j}$, produced when a distribution of photons, $\rho_{j}$, in polarisation state $\ket{j}$, collides with a focussed laser pulse. Treating the pulse as, locally, a plane wave, one integrates the plane wave NBW probability with the photon distribution, over the transverse structure of the pulse~\cite{DiPiazza:2016maj,Gelfer:2022mto}, which yields
\be
N_{j} =  \frac{2\alpha}{\eta}\int\!\mathrm{d}^{2}\mbf{x}^{\perp} \!\int_0^1\!\mathrm{d}s\,  \rho_{j}(s) \int \!\ud\vphi\, \frac{\partial\mathcal{M}_{jj}^{i}[\chi(\vphi)]}{\partial \vphi} \;,
\nonumber
\ee
where $\mbf{x}^{\perp}$ are the transverse co-ordinates, $\vphi=\vkap \cdot x$ the laser phase, and $s \in [0, 1]$ is the photon lightfront momentum fraction -- the ratio of energies $\eta/\eta_0$, for $\eta_0$ the maximum value of $\eta$. {(It is assumed that the photon distribution does not vary greatly over the focus of the laser pulse.)} Finally, $\mathcal{M}^i_{jj}$ is the imaginary part of the photon forward scattering amplitude, which is, explicitly~\cite{Baier:1975ff},
\be
\frac{\partial\mathcal{M}_{jj}}{\partial \vphi} =
\frac{2}{3} \int_{4}^{\infty}\!\frac{\mathrm{d}v}{i\pi}\, \frac{1}{z}\frac{v-4+3j}{v\sqrt{v(v-4)}}
\int \!\mathrm{d}t\,t\mbox{e}^{iz(\chi)t +it^{3}/3}
, \label{eqn:Ifunc}
\ee
for $j=1,2$ and where $z(\chi)=(v/\chi)^{2/3}$. 

Writing $\chi_{0} = \eta_{0}\xi_{0}$, where $\xi_{0}$ is the maximum value of $\xi$, so that $\chi(s,\vphi,\mbf{x}^{\perp}) = s\eta_{0}\xi(\vphi,\mbf{x}^{\perp})=\chi_{0}sf(\vphi,\mbf{x}^{\perp})$ (where $f$ describes the spacetime dependence of the laser), we see that $\chi_{0}$ is the single non-trivial input parameter for calculating the total number of pairs $N=N_{1}+N_{2}$. To make this explicit, we define $F(\chi_{0}) = N/\eta_{0}\xi_{0}^{2}$ which depends \emph{solely} on $\chi_0$. This is the function which we will Hilbert transform.
(Note that this choice of $F$ assists convergence of the numerical Hilbert transform as $F(\chi_{0}) \to 0$ quicker than $1/|\chi_{0}|$ as $\chi_{0} \to \infty$.) Then suppose, for a range of $\chi_{0}$ values, the number of pairs has been experimentally measured to produce a dataset for the pair yield, equivalently $F^{i}(\chi_0)$. This dataset is then linearly interpolated to acquire a curve $\expF^{i}$, on which one can perform a numerical Hilbert transform to obtain the estimate $\overline{F^{r}}(\chi_{0}) := {\mathcal{H}}[\expF^{i};\chi_{0}]$ of the actual real part $F^{r}(\chi_{0})$ of the elastic photon scattering process. The accuracy of $\overline{F^{r}}(\chi_{0})$ as an approximation to $F^{r}(\chi_{0})$ naturally depends on how much of the curve $F^{i}(\chi)$ is known; this means the range of $\chi$ that $F^{i}$ is measured over in experiment, as well as the accuracy of the individual measurements. We illustrate this with an idealised example before moving on the actual setup of experimental interest.

\begin{figure}[t!!]
\includegraphics[width=\linewidth]{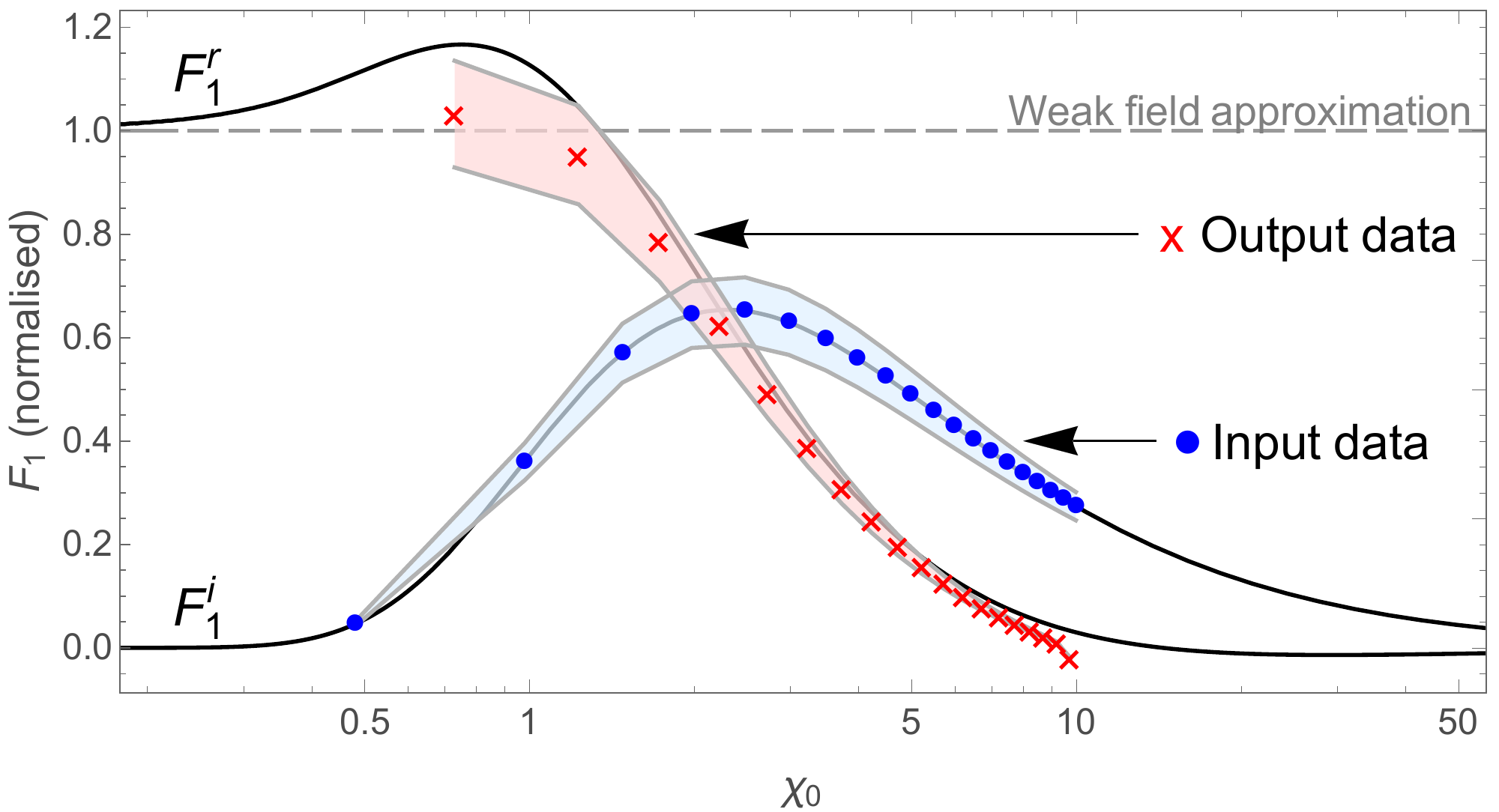}
\caption{The real and imaginary parts of $F_1$ (solid curves) normalised to its weak field limit, assuming monoenergetic, completely polarised photons propagating in a constant crossed field. (The urves for $F_2$ are similar up to an overall constant scale factor.) The blue/filled circles sample the imaginary part, representing the measured pair yield, which is 
input data for the numerical Hilbert transform. The red crosses are the output data $\overline{F^{r}_1}$, which estimates the real part $F^r_1$. The bands model a 10\% (systematic) uncertainty in the pair yield measurement. The leading-order Heisenberg-Euler approximation to $F_1$ yields the horizontal dashed line, included to emphasise that we work beyond the weak-field regime. 
\label{fig:FrANDFi}}
\end{figure}

Suppose the distribution of photons is monoenergetic and completely polarised in one state, and suppose the laser is modelled as a constant crossed field (a zero frequency plane wave). We construct the vacuum polarisation $F^{r}$ and pair yield $F^{i}$ quantities in this setup. We pick a range of $\chi_{0}$ starting at $\xi_{0}=2.5$ ($\chi_{0}=0.5$), within the region of validity of the locally constant field approximation~\cite{Blackburn:2021cuq} which we also use for our main results, below. A maximum value of  $\chi_{0}=10$ is chosen. Performing the Hilbert transform of the pair data results in the approximation $\overline{F^{r}}$ to the vacuum polarisation effect plotted in~\figref{fig:FrANDFi} {(see \Supplementary A~for details of this step)}. There is overall good agreement, even though the numerical Hilbert transform slightly underpredicts the true value of $F^{r}$. This is due to the fact that the parts of $F^{i}$ lying outside the measured region are missing from the transform -- however, their contribution is small. We also see that if there is an uncertainty in the measured pair yield on the order of 10\%, a comparable level of uncertainty is transferred to the prediction of the vacuum polarisation effect. The curves in \figref{fig:FrANDFi} are normalised to their weak-field limit (which is purely real since pair-creation is suppressed in the limit). In fact the weak-field approximations to $F$ are straight lines, hence \figref{fig:FrANDFi} emphasises that, both here and below, we go beyond the weak field approximation.

\begin{figure}[t!!]
\includegraphics[width=0.9\linewidth]{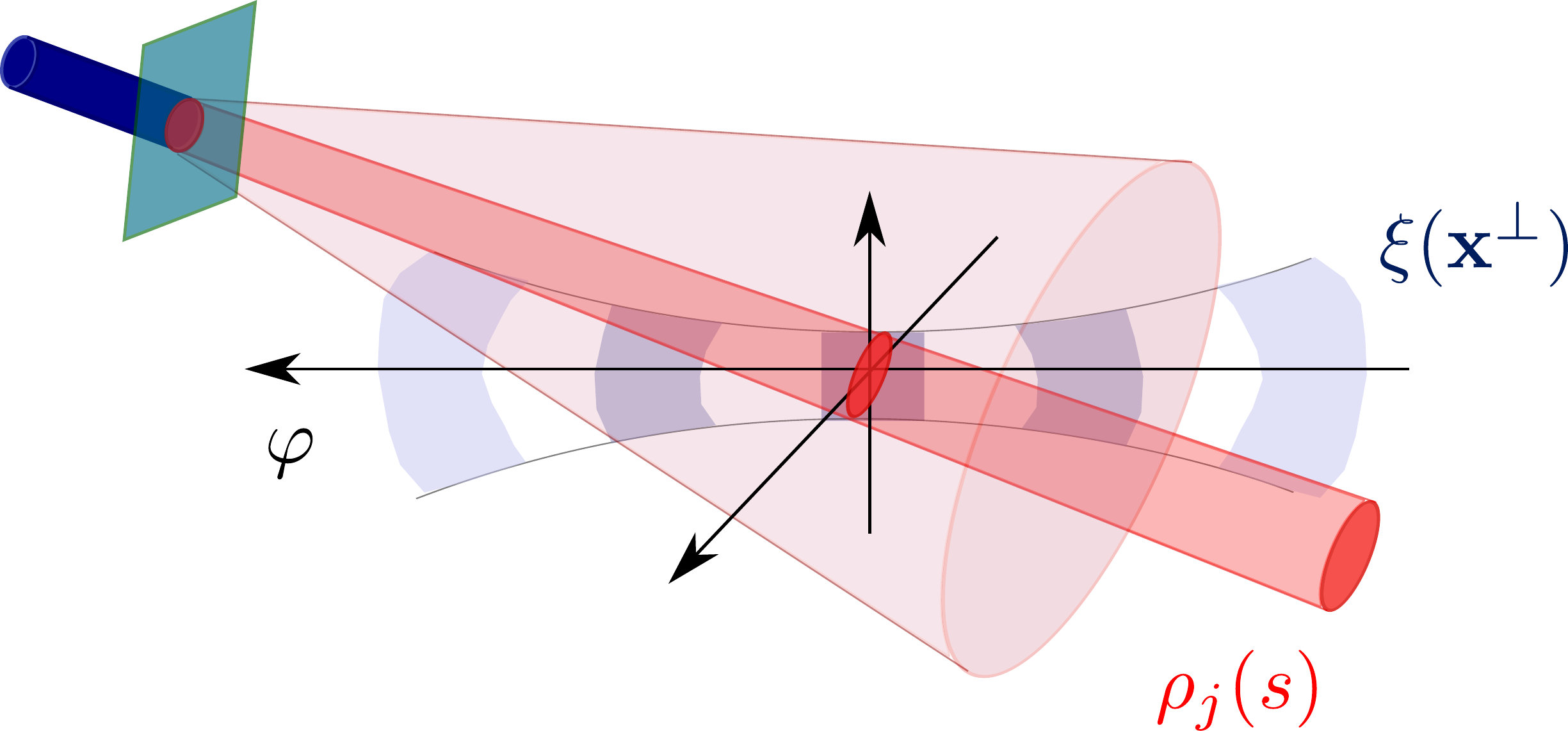}
\caption{A partially polarised source of coherent bremsstrahlung photons, $\rho_{j}(s)$ collides with a focussed laser pulse with intensity parameter $\xi(\mbf{x^{\perp}})$ and the overlap with the focal spot provides a natural collimation.\label{fig:schematic}}
\end{figure}

\emph{Setup}. To assess our scheme in a realistic case, we turn to a set-up similar to the planned LUXE experiment~\cite{Abramowicz:2021zja}. We consider a $16.5\,\trm{GeV}$ monoenergetic beam of electrons colliding with a thin target diamond radiator, which produces partially polarised coherent bremsstrahlung (CB). The bremsstrahlung photons then collide at fixed angle with a focussed laser pulse, and the overlap with the focal spot provides natural collimation of the photons. {A specific collimation is assumed in calculation of the CB spectrum; details of this, and how the yield for partially polarised photons is inverted to acquire the yield from photons in state $\ket{j}$, are given in {\Supplementary B and C respectively.}} The scenario is sketched in \figref{fig:schematic}.
We now present two example results.

\begin{figure}[t!!]
\includegraphics[width=0.99\linewidth]{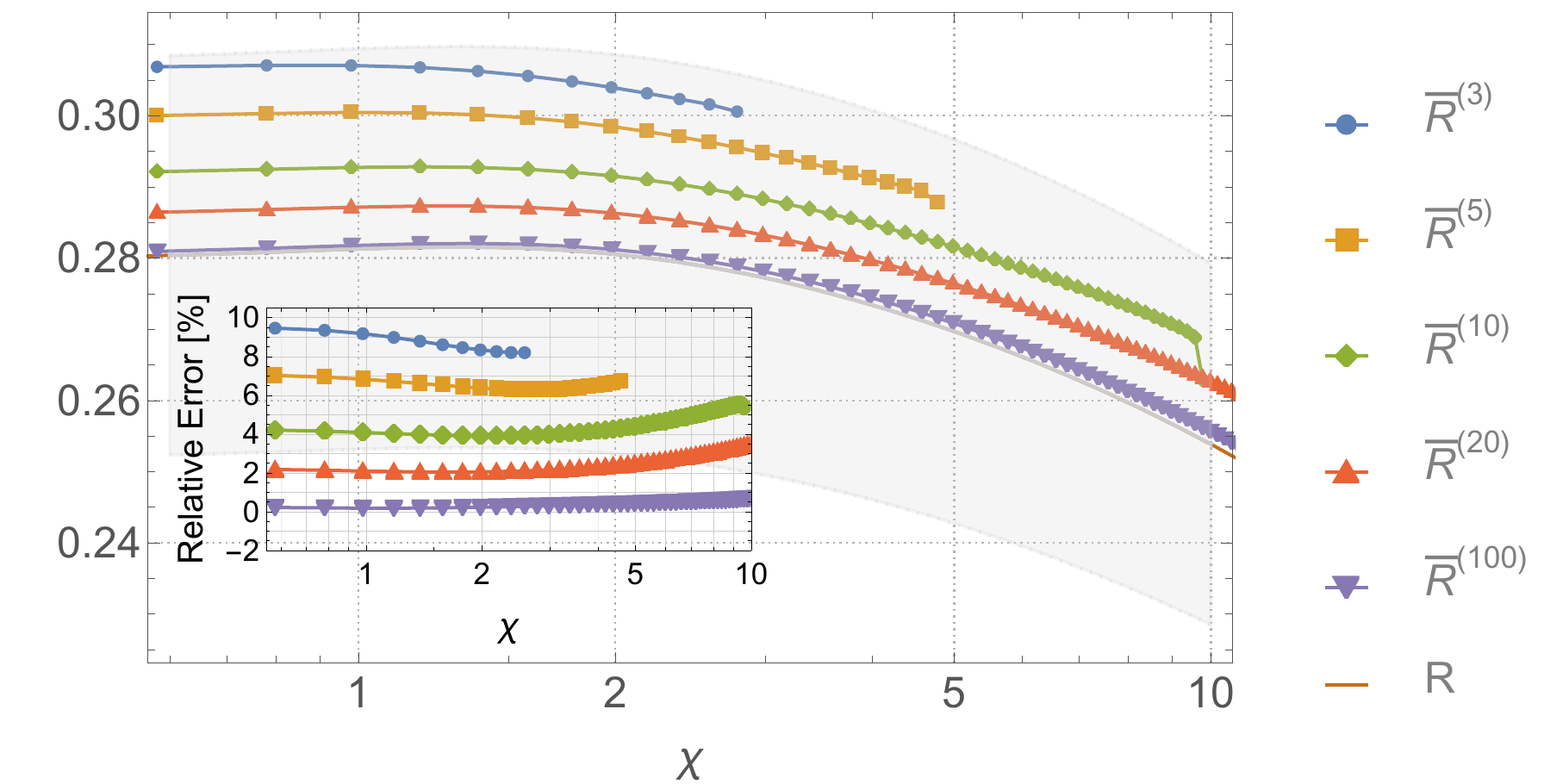}
\caption{Results for the birefringence measure $\overline{\tsf{R}}$ for the set-up described in the text. The results $\overline{\tsf{R}}^{(\chi_{\trm{max}})}$ correspond to using measurements of the number of pairs for $\chi$ values up to $\chi_{\trm{max}}$, which are to be compared to the exact value, given by $\tsf{R}$. The gray region indicates $\pm 10\%$ of the true value, and the relative error to the true value of each curve is given in the inset.} \label{fig:CBplot0}
\end{figure}

\paragraph*{Vacuum birefringence:}
%
The envisaged set-up using CB photons colliding with a focussed Gaussian laser pulse, combined with a Hilbert-transform of the measured pair yield, allows us to infer the full amplitude for photon helicity flip using~(\ref{eqn:AmpFlip}). Define the ratio of flip to {average no-flip amplitude $\tsf R = 2\mathcal{M}^r_{\pm} / (\mathcal{M}^r_{11}+\mathcal{M}^r_{22})$} in which, note, we take only the real parts. These are inferred from the experiment and Hilbert transform, which yields an estimate $\overline{\tsf{R}}$ to be compared with the theoretical {prediction} $\tsf R$. This ratio is particularly convenient to investigate as it connects directly to vacuum birefringence, which is the macroscopic result of photon helicity flip.  To see this we note that in the (low energy) regime  $\chi,\eta\ll 1$, pair creation is suppressed and the photon-scattering amplitudes are well-approximated by the replacement $\mathcal{M}_{jj}\to \mathcal{M}_{jj}^{r}$ i.e.~using just the real part. This allows one to describe vacuum polarisation effects using a semi-classical approach, based upon a real `vacuum refractive index', $n_{j}=1+\delta n_{j}$~\cite{Toll:1952rq,Baier:1967zzc}. {For photons with energy $\omega$ in polarisation eigenstate $\ket{j}$, the leading-order weak-field Heisenberg-Euler result for $\delta n_j$ is}
\be
\delta n_{j} = -\alpha m^{2}\mathcal{M}_{jj}^{r\,\prime}/\omega^{2} \simeq c_{j}\chi^{2} m^{2}/(k^{0})^{2} \;, 
\ee
in which the {well-known} low-energy constants of QED are $c_{j}=\alpha(1+3j)/90\pi$. Photons in different polarisation states thus experience different dispersion relations in the quantum vacuum, and, from (\ref{eqn:AmpFlip}), the helicity flip probability is supported on the difference of the refractive indices, which is birefringence. In the low energy, low $\chi$ regime, all volumetric factors entering ${\bar{\tsf R}}$ cancel, giving the approximation {$\tsf{R}\approx \tsf{R}^{(wf)}$ where}
\bea
{\tsf{R}^{(wf)} = \frac{c_{1}-c_{2}}{c_{1}+c_{2}}.}
\label{eqn:DeltaNwf}
\eea
{For QED, $\tsf{R}^{(wf)} = 3/11 \approx 0.273$.}
We calculate $\overline{\tsf{R}}$ for our set-up and compare with the exact value in \figref{fig:CBplot0}. The larger the maximum value of $\chi_{0}$ for which the pair yield is measured, the more accurate the prediction of vacuum birefringence. We find that for the range $0.5<\chi_{0}<3$, the prediction is accurate to within $10\%$. By comparison, in stage 1 of the LUXE experiment, $\chi_{0}$ can be varied up to around $\chi_{0}=4.5$ \cite{Abramowicz:2021zja}, which would improve the accuracy of the result. {We note} that the small-$\chi$ limit in Fig.~\ref{fig:CBplot0} compares well with the theoretical approximation~(\ref{eqn:DeltaNwf}).

\begin{figure}[t!!]
\includegraphics[width=0.99\linewidth]{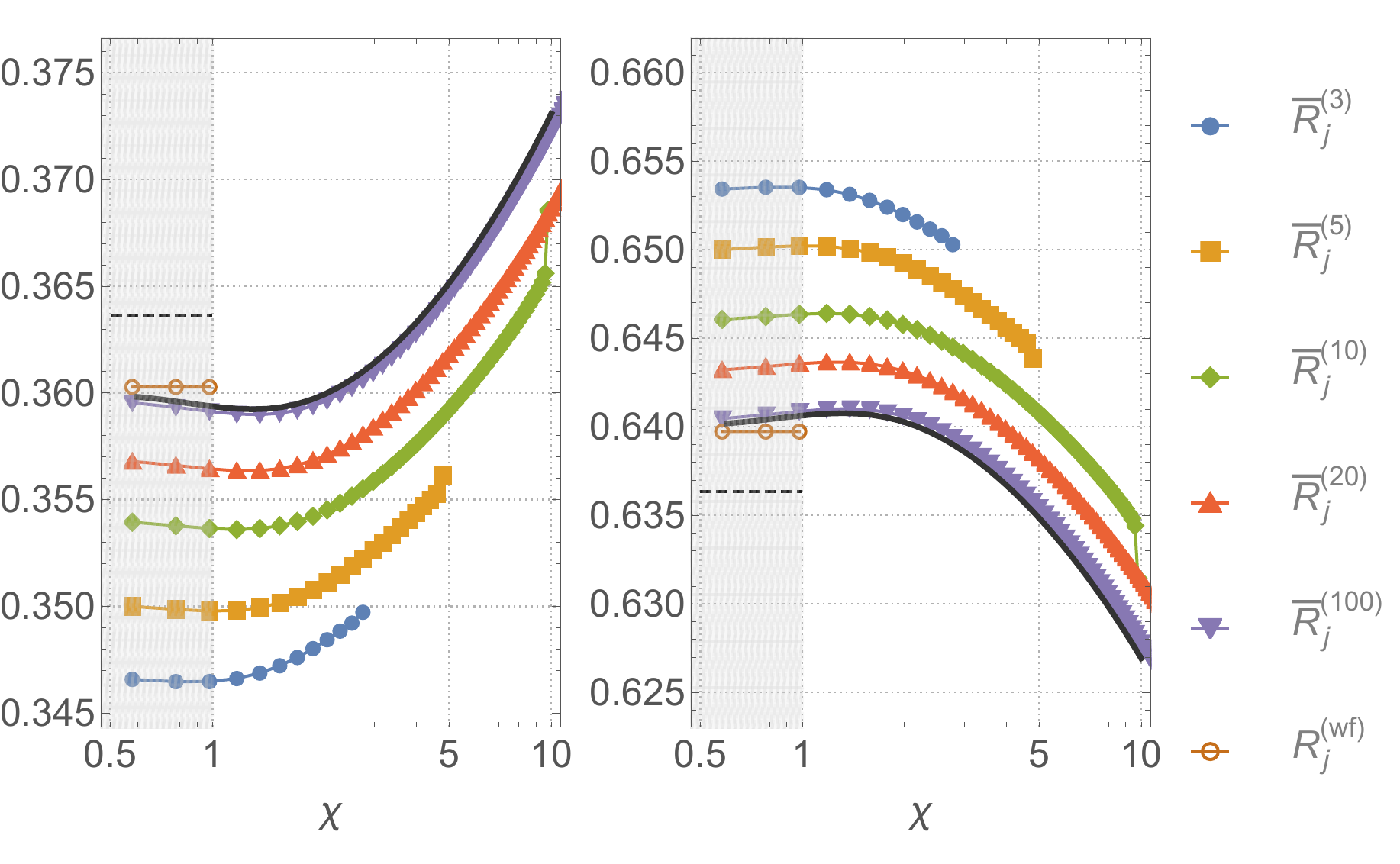}
\caption{Plot of ${\bar {\tsf R}}_j$, the ratio of `no-flip' {amplitude} in state $\ket{j}$ to average `no-flip' {amplitude}, for $j=1$ (left) and $j=2$ (right). The solid black lines are the exact results. The curves with data points are example projections from Hilbert-transforming the yield of pairs created by photons in the two polarisation eigenstates. The low-energy region of $\chi<1$ is shaded in grey and the low-energy approximation {$\tsf{R}^{(\tiny\tsf{wf})}_{j}$} (the leading-order term in a weak-field expansion of the Heisenberg-Euler result) is also shown. The black dashed line corresponds to the coarse approximation of just using the low-energy constants, i.e. the idealised situation assuming photons completely polarised in an eigenstate, giving {$\tsf{R}_{1}\approx 4/11\approx0.363$ and $\tsf{R}_{2}\approx 7/11\approx0.636$.}} \label{fig:LowEnergyConstants}
\end{figure}

\paragraph*{Low energy constants:}
%
An advantage of our approach is that pair creation from photons in a given polarisation state relates directly to `no-flip' observables (see~\eqnref{eqn:diag01}) and to the corresponding vacuum refractive index along an eigenpolarisation. Thus a second result is that our methods give us access to \emph{both} of the individual low-energy constants of QED.
This is in contrast to measuring the effects of vacuum birefringence directly, such as via the induced ellipticity in a linearly-polarised probe, which is only sensitive to the difference of the refractive indices. To see this, we calculate {$\overline{\tsf{R}}_{j} = 2\overline{\mathcal{M}}^r_{jj}/(\overline{\mathcal{M}}^r_{11}+\overline{\mathcal{M}}^r_{22})$} for our considered set-up. The results are presented in Fig.~\ref{fig:LowEnergyConstants}. We highlight, in the plot, the low-energy/weak-field approximations {${\overline{\tsf R}_{j}} \sim c_j /(c_1 + c_2)$}. It is clear both that we match these to a good approximation at low $\chi$, and that we are also sensitive to the deviation from low-energy scaling that occurs as one increases the strong-field parameter~$\chi$. (See also~ \cite{Fouche:2016qqj,Karbstein:2022uwf} for other approaches to the determination of the individual constants.)

\emph{Conclusion}.--
We have shown how photon-photon scattering effects in QED  can be indirectly measured from experimental observation of pair production in intense laser fields. Complementary to direct searches for vacuum birefringence, our method avoids signal/noise problems associated with measuring signal photons in laser backgrounds, and {it also works with} linearly polarised laser light.  Our methods give access to both vacuum refractive indices, not just the difference. A requirement that the procedure be accurate is that the pair yield must be measured over a sufficiently broad range of $\chi$. Being based, fundamentally, on unitarity and analyticity, our ideas can in principle be extended to other processes and to higher loops. Measurements of pair yield in a parameter regime where a significant fraction of pairs are `second generation' could potentially yield insight into the Ritus-Narozhny conjecture on the higher-loop behaviour of strong-field QED at high $\chi$~\cite{Fedotov:2017conjecture,Fedotov:2022ely}. We also note that for future direct searches of {new} physics using photon-photon scattering (e.g. signs of other particles `running in the loop'), a good understanding of the {Standard Model} signal will be essential, and our indirect measurement adds to the available methods by which this signal can be determined. 

\begin{acknowledgments}
\textit{BK thanks S. Boogert, O. Borysov, L. Helary and G. Sarri for useful discussions and acknowledges support from the Deutsche Forschungsgemeinschaft (DFG, German Research Foundation) under Germany’s Excellence Strategy – EXC 2121 ``Quantum Universe'' – 390833306 and from the Engineering and Physical Sciences Research Council (EPSRC), Grant No. EP/S010319/1.}
\end{acknowledgments}

\bibliography{reviewNew}

\clearpage

\appendix

\section{A. Hilbert transform of pair data} \label{sec:Hilb1}
Denote the number of pairs measured with maximum photon strong-field parameter $\chi_{0}=\xi_{0}\eta$ as $N_{j}(\chi_{0})$, as defined in~\eqnref{eqn:Nfapp}, and the corresponding value of the function to Hilbert-transform as ${F}^{i}(\chi_{0}) = {N}_{j}/\eta_{e}\xi_{0}^{2}$. Then supposing a set of $M$ measurements $\{{F}_{i}(\chi_{0,1}),\cdots,{F}_{i}(\chi_{0,M})\}$ are made, one can perform a Hilbert transform using these points for an indirect `measurement', denoted $\overline{F^{r}}(\widetilde{\chi}_{0})$, {($\widetilde{\chi}_{0} \not \in \{\chi_{0,1},\cdots,\chi_{0,M}\}$)} which encodes, e.g.~the vacuum refractive indices. The Hilbert transform of a function $F^{i}(\chi)$ is
\be
\hilb{F_{i};\chi} = \frac{1}{\pi} \mathrm{PV} \int\!\mathrm{d}\chi'\,\frac{F^{i}(\chi')}{\chi'-\chi},
\ee
{and here, $\overline{F^{r}}(\widetilde{\chi}_{0})=\mathcal{H}[F^{i};\widetilde{\chi}_{0}]$.}
Since physical values of the $\chi$ parameter always fulfil $\chi\geq 0$, the integral must be analytically continued to negative values of $\chi$. This is achieved using the prescription
\be
F_{i}(\chi) := \trm{sgn}(\chi)F_{i}(|\chi|) \;,
\label{eqn:hilb2}
\ee
which allows us to use all measured data points twice in the same numerical Hilbert transform. The total set of points used is:
\[
\{{F}_{i}(-\chi_{0,M}),\cdots {F}_{i}(-\chi_{0,1}), {F}_{i}(\chi_{0,1}),\cdots,{F}_{i}(\chi_{0,M})\}.
\]
The principal value is taken numerically by splitting the data points into a lower set: $S_{<}=\{F_{i}(-\chi_{e,M}),\cdots F_{i}(\chi_{e,j})\}$ and an upper set: $S_{>}=\{F_{i}(\chi_{e,j+1}),\cdots,F_{i}(\chi_{e,M})\}$, and integrating over both sets independently, after which the values are combined: $\mathrm{PV}\int\! \ud\chi' = \int_{S_{<}}\ud\chi'+\int_{S_{>}} \ud\chi'$. A suitable set of $\widetilde{\chi}_{0}$ values on which to evaluate the Hilbert transform is given by the mid-points of the $\chi_{0}$ values of the measurements, so $\{\widetilde{\chi}_{0,1},\cdots, \widetilde{\chi}_{0,M-1}\}$  where $\widetilde{\chi}_{0,j} = (\chi_{0,j}+\chi_{0,j+1})/2$. The $\chi_{0,j}$ values are assumed to be uniformly distributed, which helps to prevent spurious effects from arising in the numerical Hilbert transform. 

The accuracy of the transform, in comparison to the target function, can be further increased by extrapolating beyond the range of measured values, assuming an asymptotic scaling. This is achieved by taking the measured value at maximum $\chi$, i.e.~${F}_{i}(\chi_{0,M})$ and assuming that for larger $\chi_{0}$, $F_i$ follows the asympototic behaviour ${F}_{i}(\chi_{0}) \sim \chi_{0}^{-4/3}$ implied by the (locally) constant crossed field approximation. (The probability  $\prob^{\tiny\tsf{ccf}}$ for pair-creation at large $\chi$ scales as $\prob^{\tiny\tsf{ccf}}\sim\chi^{2/3}/\eta$~\cite{Nikishov:1964zza}, and $F_{i}(\chi) = \prob^{\tiny\tsf{ccf}}/\eta\xi^{2}$). In the text, we applied this asymptotic scaling from $\chi_{0,M}$ to $\chi_{0}=10^{3}$.

\section{B. Coherent Bremsstrahlung} \label{sec:CB}
Here we describe the model used for bremsstrahlung. {As pointed out in the main text, the energy spectrum and polarisation degree of coherent bremsstrahlung depend on the extent of the collimation. Here we assume an `effective collimation' to be provided by the position and dimension of the laser focal spot that the bremsstrahlung collides with. Specifically, the coherent bremsstrahlung spectrum was calculated for collision with a laser pulse of waist $w_{0}=25\,\mu\trm{m}$ and a focal spot at a distance of $3\,\trm{m}$ from the target. (These figures correspond to $\xi=1$ in stage 0 of the LUXE experiment \cite{Abramowicz:2021zja}.)} For calculations involving bremsstrahlung at different intensity parameters $\xi'$, we scale the number of photons by the ratio of areas, i.e. $(\xi/\xi')^{2}$. This is an approximation because it assumes the same bremsstrahlung spectral content, even though the collimation angle is reduced.

Although our main interest is in coherent bremsstrahlung, we also calculate the distribution of (collimated) incoherent thin-target bremsstrahlung $\rho^{(b)}$ from an amorphous target, as a test case to compare {our analytical approach with the results of full numerical} simulations in~\cite{Abramowicz:2021zja}.  The $\rho^{(b)}$ distribution  is modelled as:
\be 
\rho^{(b)}=N_{e}\frac{X}{X_{0}}\left[1 - \exp\left(-\frac{\psi_{\tiny\tsf{col.}}^{2}}{2\,\sigma_{\tiny\tsf{tot.}}^{2}}\right)\right]\left[\frac{4}{3s}(1-s)+s\right],\label{eqn:BS1}
\ee
where $N_{e}$ is the number of electrons, $X_{0}$ the radiation length, $X$ the target thickness, $\psi_{\tiny\tsf{col.}}$ the effective collimation angle and {$\sigma_{\tiny\tsf{tot.}}$ the total angular divergence of the bremsstrahlung photon pulse. The angular divergence of the bremsstrahlung photons can be written  $\sigma_{\tiny\tsf{tot.}}=(\sigma_{e}^{2}+\sigma_{0}^{2})^{1/2}$, where $\sigma_{e}$ is the electron beam angular divergence due to multiple scattering in the target, and $\sigma_{0}$ is the intrinsic divergence of bremsstrahlung photons. The intrinsic divergence of the bremsstrahlung photons is given by $\sigma_{0}= 0.52/\gamma_{e}$, where $\gamma_{e}$ is the Lorentz gamma factor of the initial $16.5\,\trm{GeV}$ electrons, which we assume does not depend on the initial photon energy.
}

{In the calculation of the coherent bremsstrahlung (CB) distribution, $\rho^{(cb)}$, we have chosen the orientation of the diamond target such as to provide the first CB peak at around $12\,\trm{GeV}$ (corresponding to a lightfront fraction of $s=0.725$). This contribution arises from the single reciprocal lattice vector {$(2,2,0)$}. The collimation factor of the coherent bremsstrahlung was assumed to have the same form as the incoherent bremsstrahlung distribution (first square bracket in \eqnref{eqn:BS1}). The coherent component of bremsstrahlung was calculated taking into account the electron beam divergence $\sigma_{e}$ and the collimation angle $\psi_{\tiny\tsf{col.}}$ using the standard method~\cite{LOHMANN1994494,KALININ1998209}}

{From a full calculation of CB from a diamond target with thickness $X=0.005\,X_{0}$ ($X_{0}=0.6\,\trm{mm}$), we find the first two peaks of the CB distribution can be approximated as:}
\be
\rho^{(cb)} = N_{e}C_{X}\left[\frac{C_{s}^{(0)}}{s}(1-s) + C_{s}^{(1)}s+\sum_{i=1}^{2}A^{(i)}\mbox{e}^{-\frac{(s-s_{i})^{2}}{2\sigma_{i}^{2}}}\right] \label{eqn:CB1}
\ee
where $C_{X}=0.000276$, $C_{s}^{(0)} = 0.105$, $C_{s}^{(1)} = 0.066$, $A^{(1)}=0.4$, $A^{(2)}=0.076$. $\sigma_{1}=0.004$, $\sigma_{2}=0.003$ with CB peaks at $s_{1}=0.72$ $s_{2} = 0.84$ corresponding to $11.92\,\trm{GeV}$ and $13.84\,\trm{GeV}$ respectively. 
{Due to a `hard' collimation ($\psi_{\tiny\tsf{col.}} \ll 1/\gamma$) the spectral width of the CB peaks becomes especially narrow, even though the conventional condition $\psi_{\tiny\tsf{col.}} >\sigma_{e}$ is not fulfilled.}

\begin{figure}[h!!]
\fbox{~\includegraphics[width=0.9\linewidth]{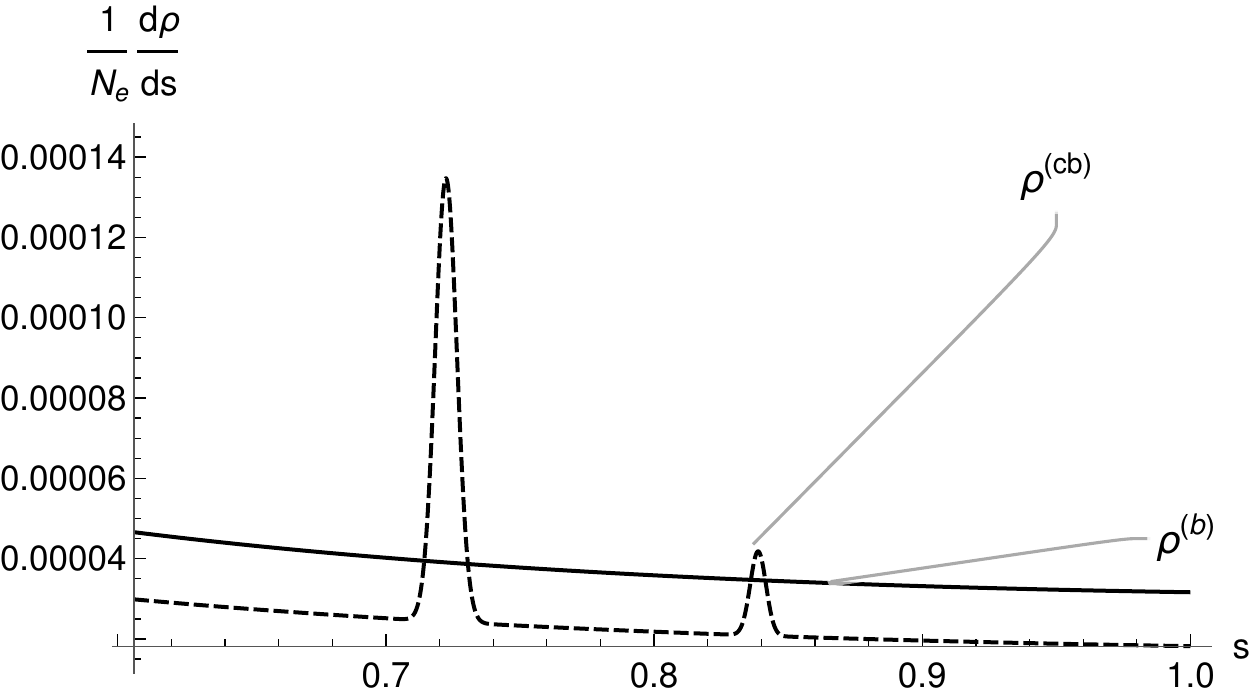}}
\caption{A comparison of the energy spectrum of the incoherent bremstrahlung photons from an amorphous target $\rho^{(b)}$, and the spectrum of coherent bremsstrahlung photons $\rho^{(cb)}$ from a diamond CVD target.} \label{fig:CBplot1}
\end{figure}

The polarisation of the CB peaks was calculated and the following parametrisation found:
\be
\Gamma_{3}= A_{1}^{\Gamma} \exp\left[-\frac{(s-s_{1})^{4}}{2(\sigma^{\Gamma}_{1})^{4}}\right]+A_{2}^{\Gamma} \exp\left[-\frac{(s-s_{2})^{4}}{2(\sigma^{\Gamma}_{2})^{4}}\right], \label{eqn:Gamma3}
\ee
where $A_{1}^{\Gamma}=0.5$, $A_{2}^{\Gamma}=0.284$, $\sigma_{1}^{\Gamma} = 0.012$, $\sigma_{2}^{\Gamma}=0.0073$, where the Stokes parameter $\Gamma_{3}$  remains practically constant in the narrow range near the CB peaks ($\Gamma_{3}=A_{n}^{\Gamma}$, for $n=1,2$) and vanishes outside. {The Stokes parameter is the asymmetry in polarisations, which we write as:
\be 
\Gamma_{3} = \frac{\rho^{(cb)}_{1}-\rho^{(cb)}_{2}}{\rho^{(cb)}_{1}+\rho^{(cb)}_{2}}.
\ee}
\begin{figure}[h!!]
\fbox{~\includegraphics[width=0.9\linewidth]{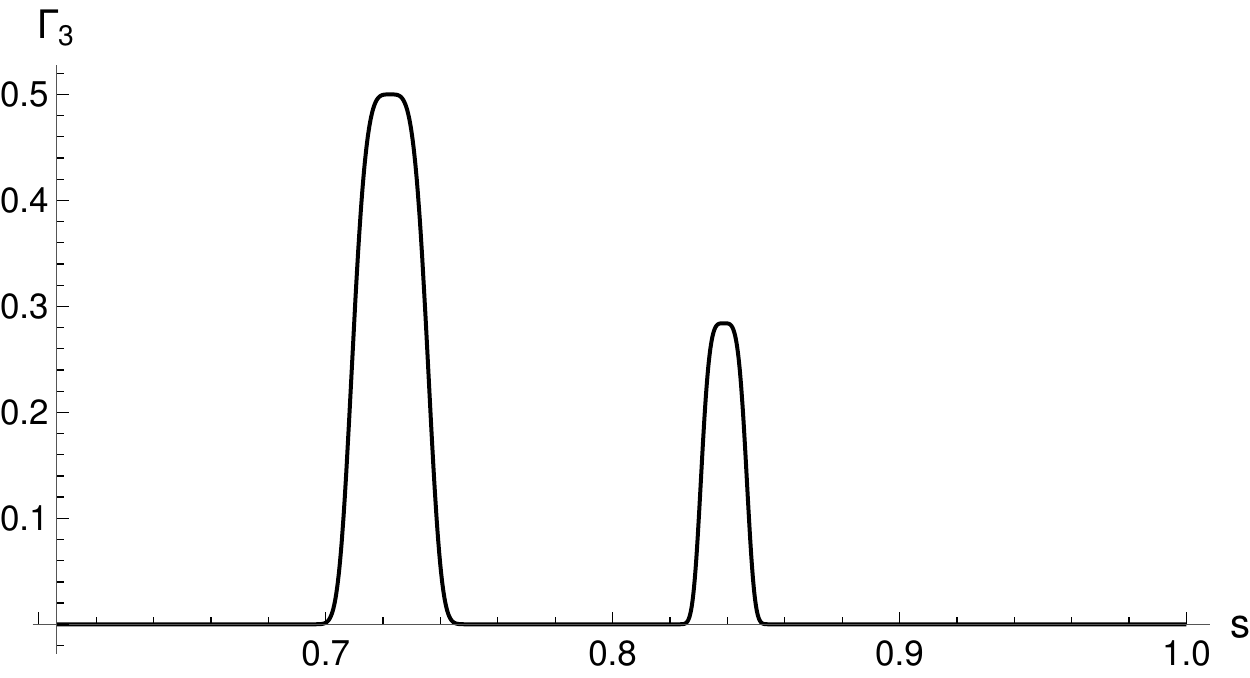}}
\caption{The Stokes parameter $\Gamma_{3}$ of the CB photons can be as large as $0.5$.} \label{fig:CBplot2}
\end{figure}
In the set-up we consider, the Stokes parameters are defined in the co-ordinate system which aligns with the crystallographic plane (2,2,0), to an accuracy of $1/\gamma$, where $\gamma$ is the Lorentz factor of the electrons generating the coherent bremsstrahlung. {(For $16.5\,\trm{GeV}$ electrons, $\gamma^{-1} \approx 3 \times 10^{-5}$.)}

In Sec.~\ref{sec:npairs}, a test of the distributions in \eqnrefs{eqn:BS1}{eqn:CB1} is made, in which the yield of pairs is calculated for LUXE parameters and compared to the simulation results. We find {good qualitative agreement and approximate quantitative agreement between the results for the full simulational approach, and the analytical results presented here}.

\section{C. Calculation of number of pairs} \label{sec:npairs}
The number of pairs is calculated using the formula
\be
N_{j} = \frac{2\alpha}{\eta}\int\! \ud^{2}\mbf{x}^{\LCperp} \int\! \ud\vphi\int_{0}^{1}\!\ud s ~\rho_{j}(s)\mathcal{M}_{jj}^{i}
[
\xi(\vphi,\mbf{x}^{\perp}),s
]. \label{eqn:Nfapp}
\ee
We model the focussed pulse using the infinite Rayleigh length approximation \cite{Gies:2017ygp,King:2018wtn} as $\xi(\vphi,\mbf{x}^{\perp})=\xi_{0}f(\vphi,\mbf{x}^{\perp})$, where:
\[
f(\vphi,\mbf{x}^{\perp}) = \mbox{e}^{-\frac{|\mbf{x}^{\perp}|^{2}}{w_{0}^{2}}}g(\vphi)\cos(\vphi),
\]
with $g(\vphi) = \partial [\sin^{2}(\vphi/2N)]/\partial \vphi$ for $0<\vphi<2\pi N$ and $N=16$ is the number of cycles, $g(\vphi)=0$ otherwise and $w_{0}$ the Gaussian waist. 

\begin{figure}[h!!]
\includegraphics[width=0.95\linewidth]{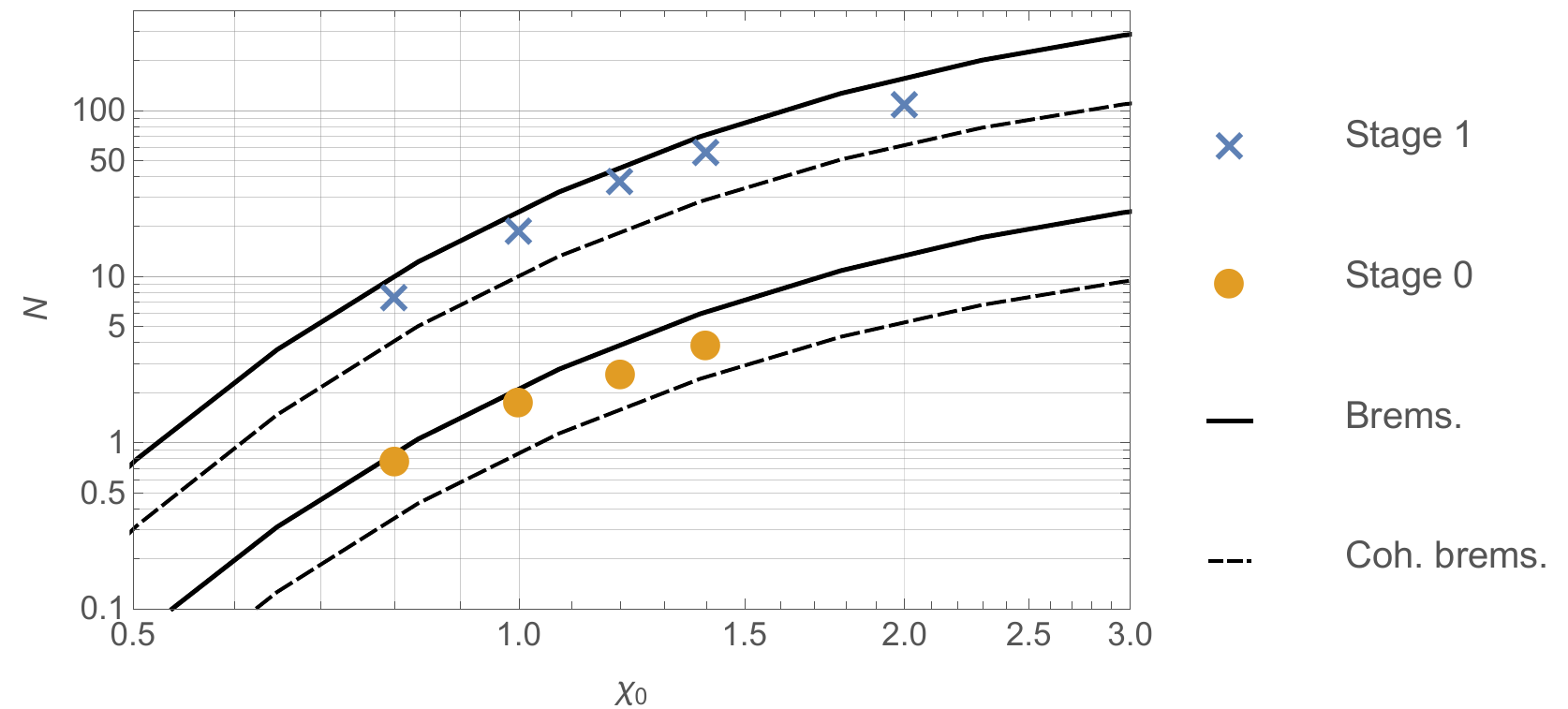}
\caption{Plot of the number of pairs calculated by \eqnref{eqn:Nfapp} using the bremsstrahlung photon distribution, $\rho^{(b)}$ from \eqnref{eqn:BS1} and (unpolarised) coherent bremsstrahlung, $\rho^{(cb)}$ from \eqnref{eqn:CB1}, compared to the simulation results from \cite{Abramowicz:2021zja} for the stage 0 and stage 1 parameters of LUXE.} \label{fig:pairs1}
\end{figure}

In \figref{fig:pairs1}, we compare the prediction of \eqnref{eqn:Nfapp} using parameters of the LUXE experiment with the results of \cite{Abramowicz:2021zja}, generated by using the code Ptarmigan \cite{Blackburn:2021cuq} to simulate the creation of pairs at the interaction point from bremsstrahlung calculated by GEANT4 \cite{Allison:2016lfl}. To compare the results in \cite{Abramowicz:2021zja}, which are for a circularly-polarised pulse, to the current work, which uses a linearly-polarised pulse, we multiply them by a factor of $2$. (This assumes that at the same $\xi$, the area of overlap between the bremsstrahlung and laser pulse can be doubled, but does not take into account any effect due to a different collimation angle.) The collimation of the bremsstrahlung and CB spectra described in Sec.~\ref{sec:CB} used parameters in stage 0 of LUXE; in order to compare with stage 1, we multiply these results by a factor equal to the ratio of powers of stage 0 and stage 1, i.e. $35/4$, to take into account a larger interaction area at constant $\xi$. (Since our aim is to verify that the order of magnitude of pairs is correct, rather than perform a high-precision comparison, also here, we do not recalculate the bremsstrahlung spectra for a new collimation angle.) As can be seen in \figref{fig:pairs1}, we find good agreement with the order of magnitude of the simulated results and our theory calculations.

In order to calculate the number of pairs from the partially polarised coherent bremsstrahlung, we must combine results from the previous and current sections. {We consider two configurations of the CB target: `$\parallel$', where the dominant CB polarisation direction is parallel to the laser polarisation and `$\perp$', where the dominant CB polarisation direction is perpendicular to the laser polarisation (both definitions assuming a head-on collision of photons with the laser pulse). The differential number of pairs produced in the two configurations is $N^{\parallel,\perp} = \int (\ud N^{\parallel,\perp}/\ud s)\, \ud s$, where}:
\bea
\frac{\ud N^{\parallel}}{\ud s} &=& \mathcal{P}(s) \frac{\ud N_{1}}{\ud s} + [1-\mathcal{P}(s)]\,\frac{\ud N_{2}}{\ud s} \nn \\
\frac{\ud N^{\perp}}{ds} &=& [1-\mathcal{P}(s)]\, \frac{\ud N_{1}}{\ud s} + \mathcal{P}(s) \frac{\ud N_{2}}{\ud s} 
\eea
where $\mathcal{P}(s) = [1+\Gamma_{3}(s)]/2$, and $\Gamma_{3}(s)$ is defined in \eqnref{eqn:Gamma3} above: the function $\mathcal{P}(s)$ then denotes the fraction of CB photons with energy fraction $s$ in polarisation state $\ket{1}$. It is then assumed that although the dependency on the photon energy of the created pairs may not be measurable in fine resolution, at least the pairs created by the first CB peak at $s=s_{1}$ can be identified, using e.g. a gamma spectrometer concept such as described in \cite{Fleck:2020opg}. Then the inversion to find the number of pairs created by photons in eigenstates $\ket{j}$, can be written as:
\bea
N_{1} &=& \frac{\mathbb{P}N^{\parallel} - (1-\mathbb{P})N^{\perp}}{2\mathbb{P}-1} \nn \\
N_{2} &=& \frac{\mathbb{P}N^{\perp}- (1-\mathbb{P})N^{\parallel}}{2\mathbb{P}-1}, 
\eea
where $\mathbb{P}$ is a constant polarisation degree that represents the polarisation of photons in the CB peak. We determined $\mathbb{P}$ by calculating the mean polarisation contributing value to pair creation. For the peak between $11.92 \pm 0.25\,\trm{GeV}$ corresponding to  $0.707<s<0.738$, this was $\mathbb{P}=0.722$ (recall from \figref{fig:CBplot2} that there is a plateau of $\Gamma_{3}=0.5$ at the centre of the peak, equivalently: $\mathcal{P}=0.75$. That this is a physically relevant choice, is evidenced by the good agreement in the low-$\chi$ limit of the results for predicted birefringence in \figref{fig:CBplot0}, with the low-$\chi$ limit of the helicity flipping amplitude.

\end{document}